\documentclass[runningheads]{llncs}

\usepackage{amsmath}
\usepackage{amssymb}
\usepackage{times}
\usepackage{float}

\floatstyle{boxed}
\newfloat{protocol}{bt}{lop}
\floatname{protocol}{Protocol}

\newtheorem{fact}{Fact}

\newcommand{\ceil}[1]   {\left\lceil #1\right\rceil}
\newcommand{\pryes}     {p_{\mathsf{yes}}}
\newcommand{\prans}[1]  {p_{\mathsf{ans}_{#1}}}
\newcommand{\prcorrect} {p_{\mathsf{ct}}}
\newcommand{\pradv} {p_{\mathsf{adv}}}
\newcommand{\Set}[1]    {\{#1\}}
\newcommand{\ot}       {\leftarrow}
\newcommand{\otr}       {\ot_R}

\newcommand{\floor}[1]  {\lfloor #1\rfloor}

\newcommand{\rket}      {\rangle}
\newcommand{\lbra}      {\langle}
\newcommand{\bra}[1]    {\langle #1|}
\newcommand{\ket}[1]    {| #1\rangle}
\newcommand{\defeq}     {:=}
\newcommand{\ZZ}        {\mathbb{Z}}

\newcommand{\var}       {\operatorname{var}}

\newcommand{\AKBOOLEAN}[1] {\mathsf{AKEncBool}(#1)}
\newcommand{\AKLIN}[1]  {\mathsf{AKLin}(#1)}

\newcommand{\COMM}[2]   {\mathsf{C}_{#1}(#2)}

\newcommand{\OT}[4] {\binom{1}{#1}\text{-}\mathsf{OT}_{#4}(#2;#3)}

\newcommand{\ADVPI}[2]     {\mathsf{Adv}^{\mathsf{pri-i}}_{#1}(#2)}
\newcommand{\ADVPR}[2]     {\mathsf{Adv}^{\mathsf{pri-r}}_{#1}(#2)}
\newcommand{\ADVCORRECT}[2]     {\mathsf{Adv}^{\mathsf{crct}}_{#1}(#2)}
\newcommand{\ADVLOR}[2]     {\mathsf{Adv}^{\mathsf{lor}}_{#1}(#2)}

\newcommand{\II}        {\mathcal{I}}
\newcommand{\RR}        {\mathcal{R}}

\newcommand{\PAR}[1]{\vspace{2mm}\noindent\textbf{#1}}

\newcommand{\comment}[1] {}

\newcommand{\infinal}[1] {#1}
\newcommand{\insubmitted}[1] {}

\newcommand{\abs}[1]    {\left| #1\right|}
\title{Cryptographic Randomized Response Techniques}

\infinal{\author{Andris Ambainis\inst{1} \and Markus Jakobsson\inst{2}
    \and Helger Lipmaa\inst{3}}

\institute{Institute of Mathematics and CS, University of Latvia, Rai\c{n}a
bulv.~29\\ R\={\i}ga, LV-1459, Latvia,\\ \email{ambainis@lanet.lv}\\ \
RSA Laboratories, 174 Middlesex Turnpike, Bedford, MA 01730, USA\\
\email{mjakobsson@rsasecurity.com}\\ \and
  Laboratory for Theoretical CS, Department of CS\&E\\
  Helsinki University of Technology, P.O.Box 5400, FIN-02015 HUT, Espoo, Finland\\
  \email{helger@tcs.hut.fi}}}

\insubmitted{\institute{} \author{*** Anonymous submission to ACM CCS
    2003 ***}}

\begin{document}

\maketitle

\begin{abstract}
  We develop cryptographically secure techniques to guarantee
  unconditional privacy for respondents to polls. Our constructions
  are efficient and practical, and are shown not to allow cheating
  respondents to affect the ``tally'' by more than their own vote ---
  which will be given the exact same weight as that of other
  respondents.  We demonstrate solutions to this problem based on both
  traditional cryptographic techniques and quantum cryptography.

\noindent
{\bf Keywords:} classical cryptography, oblivious transfer, polling,
privacy, privacy-preserving data-mining, quantum cryptography,
randomized response technique

\end{abstract}

\section{Introduction}
 
In some instances, privacy is a matter of keeping purchase information
away from telemarketers, competitors, or other intruders. In other
instances, privacy translates to security against traffic analysis,
such as for web browsing; or to security of personal location
information.  In still other instances, which we study in this paper,
privacy is a \emph{precondition} to being able to obtain answers to
important questions.  Two concrete examples of instances of latter are
\emph{elections} and \emph{surveys/polls}.

While the first of these examples is the one of the two that has
received --- by far --- the most attention in the field of
cryptography, there are important reasons to develop better privacy
tools for polling.  Surprisingly, the two examples (namely, elections
and polls), while quite similar at a first sight, are very different
in their requirements.  Since it is typically the case that there is
more funding available for providing privacy in elections than in
surveys and polls, it follows that the tallying process in the former
may involve more costly steps than that in the latter --- whether the
process is electronic (using, e.g., mix networks) or mechanic. Second,
while in the case of the voting scheme, we have that users need to
entrust their privacy with some set of authorities, it is often the
case that there is less trust established between the parties in
polls. Yet another reason to treat the two situations separately is
that elections involve many more respondents than polls typically do,
thereby allowing a unique opinion (e.g., vote) to be hidden among many
more in the case of elections than in the case of polls.  Finally,
while elections require as exact tallying as is possible,
\emph{statistical truths} are both sufficient and desirable in polls.
This allows the use of polling techniques that are very different from
election techniques --- in terms of their cost; how tallying is done;
and how privacy is protected.

While not given much attention in cryptography, important work on
polling has been done in statistics. In particular, the
\emph{randomized response technique} (RRT) was proposed by
Warner~\cite{JASA1965:Warner} in 1965, with the goal of being used in
polls relating to sensitive issues, such as drug abuse, sexual
preferences and shoplifting.  The underlying idea behind Warner's
proposal is for respondents to randomize each response according to a
certain, and known, probability distribution. More precisely, they
answer the question truthfully with some probability $\prcorrect >
1/2$, while with a fixed and known probability $1-\prcorrect$ they
lie.  Thus, users can always claim that their answer --- if it is of
the ``incriminating'' type --- was a lie. When evaluating all the
answers of the poll, these lies become statistically insignificant
given a large enough sample (where the size of the sample can be
simply computed from the probability distribution governing lying.)

However, a pure RRT by itself is not well suited for all types of
polls. E.g., it is believed that people are more likely to vote for
somebody who leads the polls than somebody who is behind.  Therefore,
it could be politically valuable not to lie (as required by the
protocol) in polls relating to ones political opinion, and therefore
have one's ``vote'' assigned a greater weight. (This is the case since
people with the opposite opinion --- if honestly following the
protocol --- will sometimes cast a vote according to your opinion, but
you would never cast a vote according to their opinion, assuming you
are willing to cheat.)  While the results of the poll remain
meaningful if \emph{everybody} cheats (i.e., tells the truth with a
probability different from that specified by the protocol), this is
\emph{not} the case when only some people deviate from the desired
behavior.  Also, while one might say that the increased weight in the
polls is gained at the price of the cheater's privacy, this is not
necessarily the case if the cheater \emph{claims} to have followed the
protocol, and there is no evidence to the contrary.

To address the problem of cheating respondents in RRT, we propose the
notion of \emph{cryptographic randomized response technique} (CRRT),
which is a modification of RRT that prevents cheating. We present
three efficient protocols for CRRT; two of them using classic
cryptographic methods (and being efficient for different values of
$\prcorrect$), and one using quantum methods.  Importantly, the
quantum RRT protocol is implementable by using contemporary
technology. We give rigorous proofs of security for one of the
classical protocols and for the quantum protocol.

For all of our proposed solutions, the privacy of the respondent will
be guaranteed information-theoretically (more precisely,
statistically).  This is appropriate to stimulate truthful feedback on
topics that may affect the respondent for years, if not decades.  All
proposed solutions also \emph{guarantee} that the respondents reply
based on the desired probability distributions.  Clearly, this
requires that the respondent cannot determine the outcome of the
protocol (as viewed by the interviewer) before the end of the
protocol.  Otherwise, he could simply halt the execution of the
protocol to suppress answers in which the communicated opinion was a
lie. We will therefore require protocols to offer privacy for the
\emph{interviewer} as well as for the respondent, meaning that the
respondent cannot learn what the outcome of the protocol is, as seen
by the interviewer. (One could relax this requirement slightly to
allow the respondent to learn the outcome at the same time as the
interviewer does, or afterward.)

While we believe that it is important to prevent the respondent from
biasing the outcome by selective halting (corresponding to the
protocol being \emph{strongly secure}), we also describe simplified
versions of our protocols in which this protection mechanism is not
available. Such simplified versions (which we refer to as \emph{weakly
  secure}) can still be useful in some situations. They may, for
example, be used as the default scheme for a given application ---
where they would be replaced by their strongly secure relatives if too
many interactions are halted prematurely. (The decision of when the
shift would be performed should be based on standard statistical
methods, and will not be covered herein.)  The benefit of considering
such dual modes is that the weakly secure versions typically are
computationally less demanding than the strongly secure versions.

Finally, we also discuss cryptographic enhancements to two alternative
RRT techniques. In the first, referred to as RRT-IQ, the respondent
always gives the truthful answer to the question he is presented with.
However, with a certain probability, he is presented with an
\underline{I}nnocous \underline{Q}uestion instead of the intended
question. A second alternative RRT technique is what is referred to as
\emph{polychotomous} RRT\@. In this version of RRT, the respondent is
given more than two possible options per question.

In particular, our first protocol uses a novel protocol for
information-theoretically secure \emph{verifiable oblivious transfer}
that enables easier zero-knowledge proofs on the properties of the
transferred values. The described protocol may also be useful in other
applications. We also note that our techniques have applications in
the privacy-preserving data-mining, see Section~\ref{rel}.

\PAR{Outline.} 
We first review the details of the randomized response technique
(Section~\ref{sec:rrt}), after which we review some related work in
cryptography (Section~\ref{rel}).  We then introduce the cryptographic
building blocks of our protocols (Section~\ref{sec:building}).  We
then describe the functionality of our desired solution in terms of
functional black boxes and protocol requirements (Section~\ref{req}).
In Section~\ref{standard}, we present our secure CRRT protocols.  In
Section~\ref{var} we describe cryptographic solutions to other
variants of the standard RRT. The appendix contains additional
information about the new oblivious transfer protocol and about the
quantum RRT protocol.

\section{Short Review of Randomized Response Technique}\label{sec:rrt}

When polling on sensitive issues like sexual behavior or tax evasion,
respondents often deny their stigmatizing behavior due to the natural
concern about their privacy. In 1965, Warner~\cite{JASA1965:Warner}
proposed the Randomized Response Technique (RRT) for organization of
polls where an unbiased estimator (UE) to the summatory information
--- the proportion of people belonging to a stigmatizing group $A$ ---
can be recovered, while the privacy of every individual respondent is
protected statistically.  Since then, different variations of the RRT
have been proposed in statistics, see~\cite{book:ChaudhuriM:1988} for
a survey. These different variations provide, for example, smaller
variance, smaller privacy breaches, optimality under different
definitions of privacy, and ability to answer polychotomous questions.
Next we will give a short overview of three types of RRT.

\PAR{RRT-W.}
In Wagner's original method (RRT-W), the respondents provide a
truthful answer to the question ``Do you belong to a stigmatizing
group $A$?'' with a certain fixed and publicly known probability
$\prcorrect > 1/2$. With probability $1-\prcorrect$ they lie --- i.e.,
answer the opposite question.  Define $\pi_A$ to be the true
proportion of the population that belongs to $A$ (or whose \emph{type}
is $t=1$).  Let $\pryes$ be the proportion of ``yes'' responses in the
poll.  Clearly, in RRT-W the \textsl{a priori} probability of getting
a ``yes'' response is $\pryes=\prcorrect \cdot
\pi_A+(1-\prcorrect)(1-\pi_A)$.  In the case of $N$ players, $L$ of
which answer ``yes'', an UE of $\pryes$ is $\widehat{\pryes}=L/N$, the
sample proportion of ``yes'' answers.  From this, one can simply
compute the unbiased estimator of $\pi_A$. This equals
$\widehat{\pi_A}=
\frac{\widehat{\pryes}-(1-\prcorrect)}{2\prcorrect-1}=
\frac{\prcorrect-1}{2\prcorrect-1}+\frac{L}{N}\cdot
\frac{1}{(2\prcorrect-1)}$. Similarly, the variance
$\var(\widehat{\pi_A})$ and its UE can be computed.

\PAR{RRT-IQ\@.} 
An alternative RRT is the~\emph{innocuous question method} (RRT-IQ),
first analyzed in~\cite{JASA1969:GreenbergASH}. When using RRT-IQ, the
respondent answers the sensitive question with a probability
$\prcorrect$, while with probability $1-\prcorrect$ to an unrelated
and innocuous question, such as ``Flip a coin.  Did you get tails?''.
The RRT-IQ achieves the same goals as RRT-W but with less
variance~\cite{book:ChaudhuriM:1988}, which makes it more suitable for
practical polling. Many other RRT-IQs are known, including some with
unknown estimate of the the proportion of the population belonging to
the innocuous group.

\comment{If an \textsl{a priori} estimate $\pi_Y$ of the proportion of
  population that belong to the innocuous group $Y$ is known, the
  respondents could be reluctant to answer truthfully. E.g., if the
  prior estimate on $\pi_Y$ is smaller than the estimate on $\pi_A$,
  answering $1$ might indicate that one belongs to the group $A$.  A
  standard method for estimating $\pi_A$ with unknown proportion
  $\pi_Y$ is to do two different polls with independent sample
  populations and different values of $\prcorrect$.  Moreover, if
  there is no protection against cheating, a respondent may be tempted
  to always answer ``no'' in order to remove any doubt regarding his
  opinion.
  
  The simplest case of RRT-IQ is one in which the unrelated question
  involves flipping a coin. As an example of RRT-IQ, we let the
  respondent answer the sensitive question with a probability
  $\prcorrect$, while he gets to flip a coin and answer whether it
  came up tails with probability $1-\prcorrect$.}

\PAR{PRRT.}
The RRTs for dichotomous polling (where the answer is yes or no) can
be generalized to \emph{polychotomous RRT} (PRRT) where the respondent
can belong to one of the $m$ mutually exclusive groups $A_1$, \ldots,
$A_m$, some of which are stigmatizing. A typical sensitive question of
this kind is ``When did you have your first child?'', with answers
``$1$ --- while not married'', ``$2$ --- within $9$ months after the
wedding'' and ``$3$ --- more than $9$ months after the wedding''. In
many cultures, the answer $1$ is stigmatizing, the answer $3$ is
innocuous, while the answer $2$ is somewhere inbetween.  The
interviewer wants to know an UE for the proportion $\pi_i$ of people
who belong to the group $A_i$, $i\in[1,m]$.  There are many possible
PRRTs~\cite[Chapter~3]{book:ChaudhuriM:1988}.  One of the simplest is
the following technique PRRT-BD by Bourke and
Dalenius~\cite{ISR1976:BourkeD}: first fix the probabilities
$\prcorrect$ and $p_1,\dots,p_m$, such that
$\prcorrect+\sum_{i\in[1,m]} p_i=1$. A respondent either reveals her
true type $t\in[1,m]$ with probability $\prcorrect$, or answers
$i\in[1,m]$ with probability $p_i$. To recover an UE of
$\vec{\pi}\defeq (\pi_1,\dots,\pi_m)^T$, define $\vec{p}\defeq
(p_1,\dots,p_m)^T$ and $\vec{\prans{}}=(\prans{1},\dots,\prans{m})^T$,
where $\prans{i}$ is the proportion of people who answer $i$. Then
$\vec{\prans{}}=\prcorrect \cdot \vec{\pi} + \vec{p}$, and hence
$\widehat{\vec{\pi}} = \prcorrect^{-1}\cdot
(\widehat{\vec{\prans{}}}-\vec{p})$.

\section{Related Cryptographic Work.}
\label{rel}

In~\cite{WIAPP1999:KikuchiANG}, Kikuchi et al.\ propose techniques
with similar goals as ours. Seemingly unaware of the previous work on
RRT, the authors reinvent this notion, and propose a protocol for
performing the data exchange.  However, their protocol is considerably
less efficient than ours. Also, it does not offer strong security in
our sense. This vulnerability to cheating makes their protocol
unsuitable for their main application (voting), as well as polls where
respondents may wish to bias their answer. Our protocols can be used
in their framework.

Our work has a relation to work on biased coin flipping, where heads
must come out with probability $\prcorrect=\ell/n$.  In our case, the
coin can be biased by the first participant in several ways, where the
choice of the distribution encodes the opinion of the respondent to
the poll. More concretely, consider a coin where one outcome (say, 1)
corresponds to ``yes'', and the other (0) to ``no''. Let us assume
that the respondent should give his correct opinion with $75\%$
probability. Then, if his opinion is ``yes'', the coin will have bias
$0.75$, while it will have bias $0.25$ if his opinion is ``no''.
However, our technique is not merely a generalization of biased coin
flipping, as we also want our protocols to implement privacy. This is
an issue that is not important in the context of ordinary biased coin
flipping.

In order to guarantee that responses are made according to the
intended distribution, we introduce a ``blinding'' requirement: we
need our protocols to be constructed such that they do not leak the
response to the respondent --- at least not until the response has
been delivered to the interviewer. From a bird's eye's view, this
makes our protocols similar to those in~\cite{CRYPTO1996:JakobssonY},
in which a party proves either language membership or language
non-membership to a verifier, but without being able to determine
which one.  However, the similarities between our protocols and those
in~\cite{CRYPTO1996:JakobssonY} do not run much deeper than that.

In contrast, there is a much closer relationship between our protocols
and protocols for oblivious transfer~\cite{Rabin81,CACM85:EvenGL}.
While our goals are orthogonal to those of oblivious transfer, the
techniques are hauntingly similar.  In particular, one of our CRRT
protocols uses a protocol for oblivious transfer as a building block.
While in principle \emph{any} such protocol can be used, it is clear
that the properties of the building block will be inherited by the
main protocol. Therefore, in order to provide unconditional guarantees
of privacy for the respondents, we use a \emph{verifiable} variant of
the information theoretic protocol for oblivious transfer, namely that
proposed by Naor and Pinkas~\cite{SODA2001:NaorP}. (An efficient
protocol that offers computational security for the sender was
proposed by Tzeng~\cite{PKC2002:Tzeng}.)

Cryptographic randomized response techniques are also related to
oblivious function evaluation~\cite{Goldreich2002:SMPC}, where one
party has data $\mu$, while another party needs to compute $f(\mu)$,
without getting to know any additional information on $\mu$, while the
first party will not get to know $f$.  Cryptographic RRTs can be seen
as protocols for oblivious function evaluation of some specific
\emph{randomized} functions $f$.

Furthermore, our work is related to the work on Private Information
Retrieval (PIR) --- and even to privacy-preserving data-mining --- in
that the goal of our interviewer is to retrieve some element from the
respondent, without the latter learning what was retrieved. More
specifically, if some $\ell$ out of $n$ elements represent the
respondent's opinion, and the remaining $n-\ell$ elements represent
the opposite opinion, then the interviewer will learn the respondent's
opinion with probability $\ell/n$ if he retrieves a random element. Of
course, in order to guarantee the interviewer that the elements are
correctly formed, additional mechanisms are required.

In privacy-preserving data-mining a related data randomization
approach has been proposed~\cite{SIGMOD2000:AgrawalS}: namely, the
users input their data to the central database (e.g., a loyal customer
inputs the name of the product he bought), and the database maintainer
needs to do some statistical analysis on the database. However, the
maintainer should not be able to recover individual items. Database
randomization in the case when the maintainer is limited to the SUM
function corresponds exactly to the RRT. For the same reasons as in
the RRT, one should not be able to bias the data. Our protocols are
also applicable in the privacy-preserving data-mining and hopefully
even in the case when more elaborated
randomizations~\cite{KDDD2002:EvfimievskiSAG} are applied.

\section{Cryptographic Building Blocks}\label{sec:building}

Assume that $p$ is a large prime, and $q$, $q\mid (p-1)$, is another
prime. Then $\ZZ_p$ has a unique subgroup $G$ of order $q$.  Let $g$
and $h$ be two generators of $G$, such that nobody knows their mutual
discrete logarithms $\log_g h$ and $\log_h g$. We let $k$ be the
security parameter, in our setting we can take $k=q$. The key $K$
consists of public parameters, $K\defeq (g;h)$.

\PAR{Pedersen's Commitment Scheme.} 
In this scheme~\cite{CRYPTO1991:Pedersen}, a message $\mu\in \ZZ_q$ is
committed by drawing a random $\rho\otr \ZZ_q$, and setting
$\COMM{K}{\mu;\rho}\defeq g^\mu h^\rho$.  The commitment can be opened
by sending $\mu$ and $\rho$ to the verifier.  This scheme is
\emph{homomorphic}, i.e.,
$\COMM{K}{\mu;\rho}\COMM{K}{\mu';\rho'}=\COMM{K}{\mu+\mu';\rho+\rho'}$.
Since it is also perfectly hiding and computationally binding, it can
be used as a building block in efficient zero-knowledge arguments,
such as protocols for arguing the knowledge of plaintext $\mu$.

\PAR{Variant of Naor-Pinkas $1$-out-of-$n$ Oblivious Transfer.}
The oblivious transfer (OT) protocol by Naor and
Pinkas~\cite{SODA2001:NaorP} guarantees information-theoretic privacy
for the sender $\RR$, and computational privacy for the chooser $\II$.
Assume the sender $\RR$ has a vector $\mu=(\mu_1,\dots,\mu_n)\in M^n$
for some set $M\subseteq \ZZ_q$. The chooser $\II$ has made a choice
$\sigma\in[1,n]$. The Naor-Pinkas protocol works as follows:

\begin{enumerate}
\item $\II$ generates random $a,b\ot \ZZ_q$ and sends $(A,B,C)\ot
  (g^a,g^b,g^{ab-\sigma+1})$ to $\RR$.
\item $\RR$ performs the following, for $i\in[1,n]$: Generate random
  $(r_i,s_i)$.  Compute $w_i\ot g^{r_i}A^{s_i}$, compute an encryption
  $y_i$ of $\mu_i$ using $v_i\ot B^{r_i}(C\cdot g^{i-1})^{s_i}$ as the
  key.  Send $(w_i, y_i)$ to $\II$.
\item $\II$ computes $w_\sigma^{b}(=v_\sigma)$ and decrypts $y_\sigma$
  using $v_\sigma$ as the key, obtaining $\mu_\sigma$.
\end{enumerate}
(Both $\RR$ and $\II$ halt if any received transcript is not correctly
formatted.)  Note that $w_i=g^{r_i+as_i}$, while $v_i=B^{r_i}(C\cdot
g^{i-1})^{s_i}=w_i^b\cdot g^{(i-\sigma)s_i}$.  Thus,
$v_\sigma=w_\sigma^b$, while for $i\neq \sigma$, $v_i$ is a random
element of $G$.  Thus, in the third step $\RR$ recovers $v_\sigma$,
while obtaining no information about $v_i$ for $i\neq \sigma$.

The Naor and Pinkas~\cite{SODA2001:NaorP} paper does not specify the
encryption method, mentioning only that the encryption scheme must be
semantically secure\comment{(e.g.,
  ElGamal~\cite{CRYPTO1984:ElGamal})}.  We propose to use Pedersen's
commitment scheme instead of an encryption scheme. Herein, we use
$K=(g;h)$ as the parameters of the commitment scheme, and use $v_i$
instead of $r_i$ as the random coin, producing a commitment $y_i\defeq
\COMM{K}{\mu_i;v_i}$.  We denote this version of Naor-Pinkas protocol,
where $y_i$ is defined as $y_i=C_K(\mu_i,v_i)$, by
$\OT{n}{\mu}{\sigma}{K}$. (The full protocol is presented in
Appendix~\ref{app:secot}.)

The idea behind this unconventional trick is that as the result, the
sender can argue in zero-knowledge for all $i\in[1,n]$ that the values
$\mu_i$ satisfy some required conditions. (We call such an OT protocol
\emph{verifiable}.) The chooser cannot decrypt $y_i$ without knowing
$v_i$, and thus he cannot guess the value of $\mu_i$ for $i\neq
\sigma$ (with probability higher than $|M|^2/q$, as we will show in
Appendix~\ref{app:secot}), even if he knows that $\mu_i$ is chosen
from a fixed two-element set. (This constitutes the security of OT
protocol in the \emph{left-or-right} sense. See
Appendix~\ref{app:secot}.) On the other hand, $\II$ can ``decrypt''
$y_\sigma$ with the ``key'' $v_\sigma$, given that the possible
message space $M$ is small enough for the exhaustive search on the set
$\Set{g^x:x\in M}$ to be practical.  In the case of dichotomous RRT,
$M=\Set{0,1}$.

\PAR{Noninteractive Zero-Knowledge Arguments.}
We will use zero-knowledge arguments (and not proofs) of knowledge in
our protocol, since they are (at the very least) statistically hiding
and computationally convincing.  This property is important in a
setting where a verifier must not be able to extract additional
information even if he is given infinite time.  \comment{A HVSZK
  argument system can be made non-interactive in the random oracle
  model by using the Fiat-Shamir heuristic~\cite{CRYPTO1986:FiatS}.}

Our first protocol uses only two very standard statistical
zero-knowledge arguments. (The arguments for the second protocol are
described in appendices.)  The first one is an argument that a given
value $y_i$ (Pedersen-)commits to a Boolean value $\mu_i\in\Set{0,1}$.
One can use standard disjunctive proofs~\cite{CRYPTO1994:CramerDS} for
this.  We denote the (possibly parallelized) argument that this holds
for $i\in[1,n]$ by $\AKBOOLEAN{y_1,\dots,y_n}$.  The second argument
of knowledge, $\AKLIN{y_1,\dots,y_{n+1};a,b}$, is
an argument that 
the prover knows some set of values $\mu_i$, for which $y_i$ is a
commitment of $\mu_i$, and such that $\sum_{i\leq
  n}\mu_i+a\mu_{n+1}=b$.  This argument of knowledge can be
constructed from Pedersen's commitment scheme by computing $y\ot
\prod_{i\leq n}y_i\cdot y_{n+1}^a$ and then arguing that the result
$y$ is a commitment to $b$. Note that such an argument of knowledge is
secure only when accompanied by zero-knowledge arguments of knowledge
of the values $\mu_i$; for this purpose, we employ
$\AKBOOLEAN{y_1,\dots,y_{n+1}}$ as described above.

\section{Security Definitions}\label{req}

In this section, we will give the definition of a weakly and strongly
secure cryptographic RRT (CRRT). The security definitions will be in
accordance with the ones in secure two-party
computation~\cite{Goldreich2002:SMPC}.  We will also explain why these
requirements are relevant in the case of CRRT\@.

Assume we have a concrete variant of RRT, like RRT-W or RRT-IQ\@. Let
$\Phi_p$ be the function that implements the desired functionality.
For example, in the case of RRT-W, $\Phi_{\prcorrect}(x)$ is a
randomized function that with probability $\prcorrect$ returns $x$,
and with probability $1-\prcorrect$ returns $1-x$.  The ideal-world
CRRT protocol, has three parties, the interviewer $\II$, the
respondent $\RR$, and the trusted third party $\mathcal{T}$. $\RR$ has
her type, $t_\RR$ as her private input, while $\II$ has no private
input. Then, $\RR$ communicates $t_\RR$ to $\mathcal{T}$, who selects
the value $r_\RR \ot \Phi_{\prcorrect}(t_\RR)$ and sends $r_\RR$ to
$\II$.  After that, the private output of $\II$ will be
$\Phi_{\prcorrect}(t_\RR)$, while $\RR$ will have no private output.
It is required that at the end of the protocol, the participants will
have no information about the private inputs and outputs of their
partners, except for what can be deduced from their own private inputs
and outputs. In particular, $\II$ (resp.  $\RR$) has no information
about the value of $t_\RR$ (resp.  $r_\RR$), except what they can
deduce from their private inputs and outputs.

In an ideal world, exactly the next three types of attacks are
possible~\cite[Section~2.1.2]{Goldreich2002:SMPC}: a party can (a)
refuse to participate in the protocol; (b) substitute his private
input to the trusted third party with a different value; or (c) abort
the protocol prematurely.  In our case, the attack (c) is irrelevant,
since $\RR$ has no output. (Attack (c) models the case when the first
party halts the protocol after receiving his private output but before
the second party has enough information to compute her output.)
Therefore, in an ideal-world RRT protocol, we cannot protect against a
participant, who (a) refuses to participate in polling
(\emph{non-participation attack}) or (b) claims that her type is
$1-t_\RR$, where $t_\RR$ is her real type (\emph{absolute denial
  attack}). No other attacks should be possible.  Note that neither
(a) nor (b) is traditionally considered an attack in the context of
polling or voting. The argument here is game-theoretic, and the
solutions must be proposed by mechanism design, instead of
cryptography: namely, a non-manipulable mechanism (e.g., the algorithm
with which the election winner is determined from all the collected
votes) must be designed so that answering against one's true type (or
non-participation) would not give more beneficial results to the
respondent than the truthful answer.

On the other hand, as we stated, no other attacks should be allowed.
This requirement is very strict, so we will explain why it is
necessary in the RRT's context.  Clearly, one must protect the privacy
of $\RR$, since this is the primarily goal of a RRT\@. It is also
necessary to protect the privacy of $\II$, although the reason here is
more subtle.  Namely, if $\RR$ obtains any additional information
about $r_\RR$ before the end of the protocol (for example, if she
suspects that $r_\RR\neq t_\RR$), she might halt the protocol.  Such a
behavior by a malicious respondent might cause a bias in the poll, as
already explained.  (Halting the protocol while having no information
on $r_\RR$ is equivalent to the non-participation attack.)  The third
requirement on the protocol, of course, is that $\II$ either halts or
receives $\Phi_{\prcorrect}(x)$, where $x$ is the input submitted by
the $\RR$.

In a real-world implementation, we want to replace $\mathcal{T}$ by a
cryptographic protocol $\Pi=(\RR,\II)$ between $\RR$ and $\II$.  This
protocol $(\RR,\II)$ is assumed to be ``indistinguishable'' from the
ideal-world protocol, that is, with a high probability, it should be
secure against all attacks that do not involve attacks (a) or (b).
``Secure'' means that the privacy of $\RR$ (resp.  $\II$) must be
protected, if $\RR$ (resp.  $\II$) follows the protocol, and that
$\II$ either halts, or receives the value $\Phi_{\prcorrect}(x)$,
where $x$ was the submitted value of $\RR$.  The security of the
respondent should be information-theoretical, while the security of
interviewer can be computational. That is, a secure CRRT-W protocol
must have the next three properties (here, $k$ is the security
parameter):

\PAR{Privacy of Respondent:}
Let $\II^*$ be an algorithm.  After the end of the protocol execution
$(\RR,\II^*)$, $\II^*$ will have no more information on $t_\RR$ than
it would have had after the execution of the ideal world protocol.
That is, assuming that $\mathsf{view}_{\II^*}$ is his view of the
protocol $(\RR,\II^*)$, define
\[
\ADVPR{k}{\RR,\II^*}\defeq \abs{\Pr[\II^*(\mathsf{view}_{\II^*},
  r_\RR)=t_\RR]- \Pr[t_\RR|r_\RR]}\enspace,
\]
where the probability is taken over the internal coin tosses of
$\II^*$ and $\RR$.  We say that a CRRT protocol is
\emph{privacy-preserving for the respondent}, if
$\ADVPR{k}{\RR,\II^*}$ is negligible (in $k$) for any unbounded
adversary $\II^*$.
  
\PAR{Privacy of Interviewer:}
Let $\RR^*$ be an algorithm.  Assume that $\II$ halts when $\RR^*$
halts. After the end of the protocol execution $(\RR^*,\II)$, $\RR^*$
will have no more information on $t_\RR$ than it would have had after
the execution of the ideal world protocol.  That is, assuming that
$\mathsf{view}_{\RR^*}$ is her view of the protocol $(\II,\RR^*)$,
define
\[
\ADVPI{k}{\RR^*,\II}\defeq
\abs{\Pr[\RR^*(\mathsf{view}_{\RR^*},t_\RR)=r_\RR]-
  \Pr[\RR^*(t_\RR)=r_\RR]}\enspace,
\]
where the probability is taken over the internal coin tosses of
$\RR^*$ and $\II$. We say that a CRRT protocol is
\emph{privacy-preserving for the interviewer}, if for any adversary
$\RR^*$, if $\ADVPI{k}{\RR^*,\II}\leq \varepsilon$ and $\RR^*$ takes
$\tau$ steps of computation then $\varepsilon\tau$ is negligible (in
$k$).
  
\PAR{Correctness:}
Let $\RR^*(x)$ be an algorithm with private input $x$ to the protocol
$(\RR^*,\II)$.  Assume that $\II$ halts when $\RR^*$ halts. We require
that at the end of the protocol execution $(\RR^*,\II)$, $\II$ will
either halt, or otherwise receive $\Phi_{\prcorrect}(x)$ with high
probability. That is, assuming that $\mathsf{view}_{\II}$ is $\II$'s
view of the protocol $(\RR^*,\II)$, define
\[
\ADVCORRECT{k}{\RR^*,\II}\defeq
1-\Pr[\II(\mathsf{view}_{\II})=\Phi_{\prcorrect}(x)| \text{$\II$ does
  not halt}]\enspace,
\]
where the probability is taken over the internal coin tosses of $\II$
and $\RR^*$.  We say that a CRRT protocol is \emph{correct}, if for
any adversary $\RR^*$, if $\ADVCORRECT {\II}{\RR^*}=\varepsilon$ and
$\RR^*$ takes up to $t$ steps of computation then $\varepsilon\tau$ is
negligible (in $k$).

We call a cryptographic RRT (CRRT) protocol \emph{weakly secure} if it
is privacy-preserving for the respondent and correct. We call CRRT
protocol \emph{(strongly) secure} if it is weakly secure and it is
privacy-preserving for the interviewer.  While a secure CRRT protocol
is preferable in many situations, there are settings where a weakly
secure CRRT protocol suffices, such as where halting can be easily
detected and punished, or means for state recovery prevent
modifications between a first and second attempt of executing the
protocol.

\section{Cryptographic RRT}\label{standard}

We will propose three different CRRT-W protocols.  In the first two
protocols, the common parameters are $\prcorrect=\ell/n > 1/2$;
generators $g$ and $h$ whose mutual discrete logs are unknown (at
least by $\RR$); and $K=(g;h)$.  $\RR$ has private input $t=t_\RR$,
and $\II$'s private output is $r_\RR$.

\PAR{CRRT Protocol Based on Oblivious Transfer.}
Our first implementation of RRT-W is described in
Protocol~\ref{prot:rrtw}.  The arguments of knowledge can be
efficiently constructed, see Sect.~\ref{sec:building}.  Here, we can
use $\AKLIN{y_1,\dots,y_{n+1};2\ell-n;\ell}$ since $\sum_{i\leq n}
\mu_i+(2\ell-n)\mu_{n+1}=\ell$ independently of the value of $t$.  All
the steps in this protocol must be authenticated.
\begin{protocol*}[ht]
  \textsc{Precomputation step:} \vspace*{-0.2cm}
  \begin{enumerate}
  \item $\RR$ prepares $n$ random bits $\mu_i\in\Set{0,1}$ for
    $i\in[1,n]$, such that $\sum \mu_i=\ell$ if $t=1$ and $\sum
    \mu_i=n-\ell$ if $t=0$.  Additionally, she sets $\mu_{n+1}\ot
    1-t$.
  \item $\II$ chooses an index $\sigma\in[1,n]$.
  \end{enumerate}
  \vspace*{-0.2cm} \textsc{Interactive step:} \vspace*{-0.2cm}
  \begin{enumerate}
  \item $\II$ and $\RR$ follow
    $\OT{n}{g^{\mu_1},\dots,g^{\mu_n}}{\sigma}{K}$. $\II$ obtains
    $g^{\mu_\sigma}$, and computes $\mu_\sigma$ from that.
  \item $\RR$ sends to $\II$ noninteractive zero-knowledge arguments
    $\AKBOOLEAN{y_1,\dots,y_{n+1}}$, and
    $\AKLIN{y_1,\dots,y_{n+1};2\ell-n;\ell}$.
  \item $\II$ verifies the arguments, and halts if the verification
    fails.
  \end{enumerate}
  \caption{A secure CRRT-W protocol based on oblivious transfer}
  \label{prot:rrtw}
\end{protocol*}

If we take the number of bits that must be committed as the efficiency
measure (communication complexity of the protocol), then our protocol
has complexity $O(n)$.  In the polling application, one can most
probably assume that $n\leq 5$.  The security proofs of this protocol
follow directly from the properties of underlying primitives. As a
direct corollary from Theorem~\ref{thm:ot}, we get that
Protocol~\ref{prot:rrtw} is privacy-preserving for respondent
($\ADVPR{k}{\RR,\II^*}\leq 2/q+O(1/q)$, where the constant comes in
from the use of statistically-hiding zero-knowledge arguments). It is
privacy preserving for interviewer, given the Decisional
Diffie-Hellman (DDH) assumption.  The correctness of this protocol
follows from the properties of the zero-knowledge arguments used under
the DDH assumption.
  
In a simplified weakly secure protocol based on the same idea, $\RR$
commits to all $\mu_i$ by computing and publishing $y_i\ot
\COMM{K}{\mu_i;\rho_i}$.  Next, $\RR$ argues that
$\AKBOOLEAN{y_1,\dots,y_{n+1}}$, and
$\AKLIN{y_1,\dots,y_{n+1};2\ell-n;\ell}$. After that, $\II$ sends
$\sigma$ to $\RR$, who then reveals $\mu_\sigma$ and $\rho_\sigma$.
Upon obtaining these, $\II$ verifies the correctness of the previous
corresponding commitment, outputting $\mu_\sigma$.

\PAR{CRRT from Coin-Flipping.}
Protocol~\ref{prot:rrtw2} depicts a secure CRRT-W protocol with
communication complexity $\Theta(d\log_2 n)$, where $d\defeq
\ceil{1/(1-\prcorrect)}$, and $\prcorrect=\ell/n$ as previously. While
in the common RRT application one can usually assume that $n$ is
relatively small, this second protocol is useful in some specific
game-theoretic applications where for the best outcome, $\prcorrect$
must have a very specific value. The idea behind this protocol is that
at least one of the integers $\mu+\nu+i\ell\mod{n}$ must be in
interval $[0,\ell-1]$, and at least one of them must be in interval
$[\ell,n-1]$.  Hence, $\II$ gets necessary proofs for both the $0$ and
the $1$ answer, which is sufficient for his goal. For his choice to be
accepted, he must accompany the corresponding $r$ with $\RR$-s
signature on his commitment on $\sigma$.

\begin{protocol*}[ht]
  \textsc{Precomputation step:} \vspace*{-0.2cm}
  \begin{enumerate}
  \item $\RR$ chooses a random $\mu\otr[0,n-1]$.
  \item $\II$ chooses random $\nu\otr[0,n-1]$ and $\sigma\otr[0,d-1]$.
  \end{enumerate}
  \vspace*{-0.2cm} \textsc{Interactive step:} \vspace*{-0.2cm}
  \begin{enumerate}
  \item $\RR$ commits to $t$ and $\mu$, and sends the commitments to
    $\II$.
  \item $\II$ commits to $\sigma$, by setting $y\ot
    \COMM{K}{\sigma;\rho}$ for some random $\rho$.  He sends $\nu$ and
    $y$ to $\RR$, together with a zero-knowledge argument that $y$ is
    a commitment of some $i\in[0,d-1]$.
  \item\label{it:exmu} $\RR$ verifies the argument. She computes
    values $\mu'_i$, for $i\in[0,d-1]$, such that $\mu'_i=t \iff
    (\mu+\nu+i\ell\mod{n})< \ell$. She signs $y$, and sends her
    signature together with $\Set{\mu'_i}$ and the next zero-knowledge
    argument for every $i\in[0,d-1]$: $[\mu'_i=t \iff
    (\mu+\nu+i\ell\mod{n})< \ell]$.
  \item After that, $\II$ sets $r_\RR\ot \mu'_\sigma$. He will
    accompany this with $\RR$-s signature on the commitment, so that
    both $\RR$ and third parties can verify it.
  \end{enumerate}
  \caption{A secure CRRT-W protocol based on coin-flipping}
  \label{prot:rrtw2}
\end{protocol*}

A weakly secure version of this protocol is especially efficient.
There, one should set $d\ot 1$, and omit the steps in
Protocol~\ref{prot:rrtw2} that depend on $\sigma$ being greater than
$1$. (E.g., there is no need to commit to $\sigma$ anymore.) Thus,
such a protocol would have communication complexity $\Theta(\log_2
n)$.  Now, $\prcorrect> 1/2$ (otherwise one could just do a bit-flip
on the answers), and hence $d>2$. On the other hand, the privacy of
respondents is in danger if say $\prcorrect\geq 3/4$. Thus, we may
assume that $d\in[3,4]$. Therefore, Protocol~\ref{prot:rrtw2} will be
more communication-efficient than Protocol~\ref{prot:rrtw} as soon as
$n/\log_2 n>4\geq d$, or $n\geq 16$. The weakly secure version will be
\emph{always} more communication-efficient.

This protocol is especially efficient if the used commitment scheme is
an integer commitment
scheme~\cite{IEICE1999:FujisakiO,ASIACRYPT2002:DamgaardF}. In this
case, to argue that $(\mu+\nu+i\ell\mod{n})<\ell$ one only must do the
next two simple steps: first, argue that $\mu+\nu+i\ell=z+e n$ for
some $z$, $e$, and then, argue that $z\in[0,\ell-1]$. This can be done
efficiently by using the range proofs
from~\cite{EUROCRYPT2000:Boudot,Lipmaa:szkde:2001}. One can also use
Pedersen's scheme, but this would result in more complicated
arguments. \comment{See~\cite{EUROCRYPT2000:Boudot} for references.}

\PAR{Quantum-Cryptographic RRT.}
We also present a \emph{quantum CRRT protocol} (see
Protocol~\ref{prot:qrrt}) that allows for a value $\prcorrect$ that
does not have to be a rational number, and which provides a relaxed
form of information-theoretic security to \emph{both} parties.  While
not secure by our previous definitions, it provides meaningfully low
bounds on the probabilities of success for a cheater.  Namely, (a) if
dishonest, $\RR$ cannot make his vote count as more than $\sqrt{2}$
votes: if $\prcorrect=\frac{1}{2}+\varepsilon$, then $\pradv\leq
\frac{1}{2}+\sqrt{2}\varepsilon$ (we also show a slightly better bound
with a more complicated expression for $\pradv$, cf. Appendix
\ref{quantumdetails}).  (b) if dishonest strategy allows $\II$ to
learn $t$ with probability $\prcorrect+\varepsilon$, it also leads to
$\II$ being caught cheating with probability at least
$\frac{2\prcorrect-1}{2}\varepsilon$.
This form of security (information-theoretic security with relaxed
definitions) is common for quantum protocols for tasks like bit
commitment \cite{STOC2000:AharonovTVY} or coin flipping
\cite{STOC2001:Ambainis,PRL2002:SpekkensR}. The security guarantees of
our quantum protocol compare quite well to ones achieved for those
tasks.  A desirable property of this quantum protocol is that it can
be implemented by using contemporary technology, since it only
involves transmitting and measuring single qubits, and no maintaining
of coherent multi-qubit states.

\begin{protocol*}[tp]
  \textsc{Precomputation step:} \vspace*{-0.2cm}
  \begin{enumerate}
  \item $\II$ chooses random $u_0\otr[0,1]$, $u_1\otr[0,1]$.  He
    generates quantum states
    $\ket{\psi_0}=\sqrt{\prcorrect}\ket{u_0}+\sqrt{1-\prcorrect}\ket{1-u_0}$,
    $\ket{\psi_1}=\sqrt{\prcorrect}\ket{u_1}+\sqrt{1-\prcorrect}\ket{1-u_1}$.
  \item $\RR$ chooses a random $i\otr[0,1]$.
  \end{enumerate}
  \vspace*{-0.2cm} \textsc{Interactive step:} \vspace*{-0.2cm}
  \begin{enumerate}
  \item $\II$ sends $\ket{\psi_0}$ and $\ket{\psi_1}$ to $\RR$.
  \item $\RR$ sends $i$ to $\II$.
  \item $\II$ sends $u_i$ to $\RR$.
  \item $\RR$ measures the state $\ket{\psi_i}$ in the basis
    $\ket{\psi_{u_i}}=\sqrt{\prcorrect}\ket{u_i}+
    \sqrt{1-\prcorrect}\ket{1-u_i}$,
    $\ket{\psi^{\perp}_{u_i}}=\sqrt{1-\prcorrect}\ket{u_i}-
    \sqrt{\prcorrect}\ket{1-u_i}$ and halts if the result is not
    $\ket{\psi_{u_i}}$.
  \item If the verification is passed, $\RR$ performs the
    transformation $\ket{0}\rightarrow \ket{t}$, $\ket{1}\rightarrow
    \ket{1-t}$ on the state $\ket{\psi_{1-i}}$ and sends it back to
    $\II$.
  \item $\II$ measures the state in the basis $\ket{0}$, $\ket{1}$,
    gets outcome $s$. $\II$ outputs $r\ot u_i\oplus s$.
  \end{enumerate}
  \caption{A quantum CRRT-W protocol.}
  \label{prot:qrrt}
\end{protocol*}

To show the main ideas behind quantum protocol, we now show how to
analyze a simplified version of protocol \ref{prot:qrrt}.  The
security proof for the full protocol is quite complicates and is given
in appendix \ref{quantumdetails}.  We also refer to appendix
\ref{quantumdetails} for definitions of quantum states and operations
on them.

The simplified version of Protocol~\ref{prot:qrrt} is:
\begin{enumerate}
\item $\II$ chooses a random $u\otr[0,1]$, prepares a quantum bit in
  the state
  $\ket{\psi_u}=\sqrt{\prcorrect}\ket{u}+\sqrt{1-\prcorrect}\ket{1-u}$
  and sends it to $\RR$.
\item $\RR$ performs a bit flip if her type $t=1$, and sends the
  quantum bit back to $\II$.
\item $\II$ measures the state in the computational basis $\ket{0}$,
  $\ket{1}$, gets answer $s$. The answer is $r=u\oplus s$.
\end{enumerate}

If both parties are honest, the state returned by respondent is
unchanged: $\sqrt{\prcorrect}\ket{u}+\sqrt{1-\prcorrect}\ket{1-u}$ if
$t=0$ and $\sqrt{\prcorrect}\ket{1-u}+\sqrt{1-\prcorrect}\ket{u}$ if
$t=1$.  Measuring this state gives the correct answer with probability
$1-\prcorrect$.  Next, we show that respondent is unable to misuse
this protocol.

\begin{theorem}\label{thm:q-honesti}
  For any respondent's strategy $\RR^*$, the probability of honest
  interviewer $\II$ getting $r=1$ is between $1-\prcorrect$ and
  $\prcorrect$. Therefore, the previous protocol is both correct and
  privacy-preserving for the interviewer.
\end{theorem}
\begin{proof}
  We show that the probability of $r=1$ is at most $\prcorrect$.  The
  other direction is similar.  We first modify the (simplified)
  protocol by making $\RR^*$ to measure the state and send the
  measured result to $\II$, this does not change the result of the
  honest protocol since the measurement remains the same.  Also, any
  cheating strategy for $\RR^*$ in the original protocol can be used
  in the new protocol as well.  So, it is sufficient to bound the
  probability of $r=1$ in the new protocol.
  
  Now, the answer is $r=1$ if $\II$ sent $\ket{\psi_i}$ and $\RR^*$
  sends back $j$, with $i=j$.  Thus, we have the setting of
  Fact~\ref{TTheoremPure} (see Appendix~\ref{sec:qback}).  The rest is
  a calculation: to determine the angle $\beta$ between $\ket{\psi_0}$
  and $\ket{\psi_1}$, it suffices to determine the inner product which
  is $\sin\beta=2\sqrt{\prcorrect(1-\prcorrect)}$.  Therefore,
  $\cos\beta=\sqrt{1-\sin^2 \beta}=2\prcorrect-1$ and
  $\frac{1}{2}+\frac{\cos\beta}{2}=\prcorrect$.  \qed
\end{proof}

On the other hand, when using this simplified version, a dishonest
interviewer $\II^*$ can always learn $t$ with probability 1.  Namely,
it suffices to send the state $\ket{0}$.  If $t=0$, $\RR$ sends
$\ket{0}$ back unchanged.  If $t=1$, $\RR$ applies a bit flip. The
state becomes $\ket{1}$.  $\II$ can then distinguish $\ket{0}$ from
$\ket{1}$ with certainty by a measurement in the computational basis.

Note that this is similar to a classical ``protocol'', where $\II$
first generates a random $u$ and sends a bit $i$ that is equal to $u$
with probability $\prcorrect$ and $1-u$ with probability
$1-\prcorrect$.  $\RR$ then flips the bit if $t=1$ and sends it back
unchanged if $t=0$. The interviewer XORs it with $u$, getting $t$ with
probability $\prcorrect$ and $1-t$ with probability $1-\prcorrect$.
In this "protocol", $\RR$ can never cheat.  However, $\II^*$ can learn
$t$ with probability $1$ by just remembering $i$ and XORing the answer
with $i$ instead of $u$.  In the classical world, this flaw is fatal
because $\II$ cannot prove that he has generated $i$ from the correct
probability distribution and has not kept a copy of $i$ for himself.
In the quantum case, $\II$ can prove to $\RR$ that he has correctly
prepared the quantum state.  Then, we get Protocol~\ref{prot:qrrt}
with $\II$ sending two states $\ket{\psi_{u_0}}$ and
$\ket{\psi_{u_1}}$, one of which is verified and the other is used for
transmitting $t$.  (See Appendix \ref{quantumdetails} for detailed
analysis of this protocol.)

\section{Protocols for Other RRTs and Extensions}
\label{var}

\PAR{Protocol for Cryptographic RRT-IQ\@.}
Recall that in one version of RRT-IQ, the respondent would reply with
his true opinion $t_\RR$ with a rational probability
$\prcorrect=\ell/n$, while he would otherwise flip a coin and answer
whether it came up tails. Like for CRRT-W, it is important to
guarantee the use of correct distributions.  Protocol~\ref{prot:rrtw}
can be easily changed to work for this version of RRT-IQ\@. Instead of
$n$ random bits, $\RR$ prepares $2n$ random bits $\mu_i$, so that
$\sum \mu_i=n+\ell$ if $t_\RR=1$, and $\sum \mu_i=n-\ell$ if
$t_\RR=0$. She also prepares a checksum bit $\mu_{2n+1}=1-t_\RR$. The
rest of the protocol is principally the same as in
Protocol~\ref{prot:rrtw}, with $n$ changed to $2n$, and $\RR$ arguing
that $\AKLIN{y_1,\dots,y_{2n+1};2\ell;2n-\ell}$.

\PAR{Protocol for Cryptographic PRRT-BD.}
The next protocol is a modification of Protocol~\ref{prot:rrtw} as
well.  Let $p_i$ be such that $\prcorrect+\sum_{i\in[1,m]} p_i=1$, and
assume that every respondent has a type $t_\RR\in[1,m]$.  Assume
$\prcorrect=\ell/n$, $p_i=\ell_i/n$ and that $p_i=0$ if
$i\not\in[1,m]$.  Assume $D\geq \max(\ell,\ell_1,\dots,\ell_m)+1$. The
respondent prepares $n$ numbers $D^{\mu_i}$, such that
$\sharp\Set{i:\mu_i=t_\RR}=\ell_{t_\RR}+\ell$, and
$\sharp\Set{i:\mu_i=j}=\ell_j$, if $j\neq t_\RR$. Then the interviewer
and respondent will execute a variant of OT with choice $\sigma$,
during which the interviewer only gets to know the value $\mu_\sigma$.
Then the respondent argues that the sum of all commitments is a
commitment to the value $\sum \ell_i D^{\mu_i}+\ell D^j$, for some
$j\in[1,m]$, by using range-proofs in
exponents~\cite{FC2002:LipmaaAN}. (A more efficient proof methodology
is available when $D$ is a prime~\cite{FC2002:LipmaaAN}, given that
one uses an integer commitment scheme.)  Additionally, she argues that
every single commitment corresponds to a value $D^i$ for $i\in[1,m]$,
also using range-proofs of exponents~\cite{FC2002:LipmaaAN}. After the
OT step, the interviewer gets $g^{\mu_\sigma}$, and recovers
$\mu_\sigma$ from it efficiently. (Note that $m\leq 10$ is typical in
the context of polling.)

\PAR{Extensions to Hierarchies of Interviewers.}
One can consider a hierarchy of interviewers, reporting to some
central authority. If there is a trust relationship between these two
types of parties, no changes to our protocol would be required.
However, if the central authority would like to be able to avoid
having to trust interviewers, the following modifications could be
performed.  First, each respondent would have to authenticate the
transcript he generates, whether with a standard signature scheme, a
group signature scheme, etc.  Second, and in order to prevent
collusions between interviewers and respondents, the interviewers must
not be allowed to know the choice $\sigma$ made in a particular
interview. Thus, the triple $(A,B,C)$ normally generated by the
interviewer during the Naor-Pinkas OT protocol would instead have to
be generated by the central authority, and kept secret by the same.
More efficient versions of \emph{proxy} OT satisfying our other
requirements are beneficial for this
application~\cite{ASIACRYPT2000:NaorP}.

\infinal{\subsection*{Acknowledgments}
  
  We would like to thank Jouni K. Sepp\"{a}nen for introducing us to
  the RRT, and for fruitful discussions on the topic. We would like to
  thank Benny Pinkas for comments.}

\bibliographystyle{alpha}


\appendix

\section{Security of Modified Oblivious Transfer Protocol}\label{app:secot}

From our oblivious transfer protocol $\OT{n}{\mu}{\sigma}{K}$ we will
require that it must be secure in the next sense. The attack scenario
consists of the following game.  The chooser $\II^*$ chooses $\sigma$
and two different vectors, $\mu[1]=(\mu[1]_1,\dots,\mu[1]_n)\in M^n$
and $\mu[2]=(\mu[1]_1,\dots,\mu[1]_n)\in M^n$, such that
$\mu[1]_\sigma=\mu[2]_\sigma$. Denote an $\II^*$ that has made such
choices by $\II^*(\mu[1],\mu[2])$. He submits both tuples to the
responder, who flips a fair coin $b\otr[1,2]$. After that, the chooser
and the responder execute the protocol $\OT{n}{\mu[b]}{\sigma}{K}$.
After receiving $\mu[b]_\sigma$, $\II^*$ guesses the value of $b$. Let
$\ADVLOR{k}{\II^*,\RR}$ be the probability that $\II^*$ guesses the
correct $b$, where probability is taken over the internal coin tosses
of $\II^*$ and $\RR$. We say that the oblivious transfer protocol is
$\varepsilon$-secure in the \emph{left-or-right} sense, if for any
unbounded algorithm $\II^*$, $\ADVLOR{k}{\II^*,\RR}\leq \varepsilon$.

Recall that the proposed variant of the Naor-Pinkas protocol works as
follows:
\begin{enumerate}
\item $\II$ generates random $a,b\ot \ZZ_q$ and sends $(A,B,C)\ot
  (g^a,g^b,g^{ab-\sigma+1})$ to $\RR$.
\item $\RR$ performs the following, for $i\in[1,n]$: Generate random
  $(r_i,s_i)$.  Compute $w_i\ot g^{r_i}A^{s_i}$, compute an encryption
  $y_i\ot g^{\mu_i}h^{v_i}$, where $v_i\ot B^{r_i}(C\cdot
  g^{i-1})^{s_i}$.  Send $(w_i, y_i)$ to $\II$.
\item $\II$ computes $w_\sigma^{b}(=v_\sigma)$ and recovers
  $g^{\mu_\sigma}\ot y_\sigma/h^{w_\sigma^b}$.
\end{enumerate}

\begin{theorem}\label{thm:ot}
  Let $\OT{n}{\cdot}{\cdot}{K}$ be the described oblivious transfer
  protocol. (a) If a malicious $\RR^*$ can guess the value of $\sigma$
  with advantage $\varepsilon$, then he can solve the Decisional
  Diffie Hellman (DDH) problem with the same probability and in
  approximately the same time. (v) This protocol is $(m-d)(m-1)/q\leq
  m(m-1)/q$-secure in the left-or-right sense, where $d\defeq
  q\mod{m}$ and $m\defeq |M|$.
\end{theorem}

\begin{proof}[Sketch.]
  (a) Assume that $\RR^*$ can guess $\sigma$ with probability
  $\varepsilon$, given her view $(A, B, C) = (g^a, g^b, g^{a b -
    \sigma + 1}$). But then she can solve the DDH problem (given
  $(g^a, g^b, g^c)$ for random $a$ and $b$, decide whether $c = a b$
  or not) with probability $\varepsilon$: given an input $(g^a, g^b,
  g^c)$, she just computes such a $\sigma$, for which $c = a b -
  \sigma + 1$.  After that, she only has to check whether $\sigma=1$
  or not.
  
  (b) W.l.o.g., assume that $\sigma=1$. Define $\nu[j]$ to be a
  vector, for which which $\nu[j]_i=\mu[1]_i$ if $i>j$, and
  $\nu[j]_i=\mu[2]_i$ if $i\leq j$. Thus $\nu[1]=\mu[1]$ (since
  $\mu[1]_1=\mu[2]_1$), while $\nu[n]=\mu[2]$, and for all $j$,
  $\nu[j-1]$ and $\nu[j]$ differ only in the $j$th element
  $\nu[j]_j\neq \nu[j+1]_j$. Thus, our goal is to show that
  $\II^*(\nu[1],\nu[n])\leq m(m-1)/q$. For this we will prove that
  $\II^*(\nu[j-1],\nu[j])\leq (m-d)/q \leq m/q$ for every $j\in[2,n]$
  and then use the triangle equality to establish that
  $\ADVLOR{k}{\II^*(\mu[1],\mu[2]),\RR}\leq \sum_{i=2^n}
  \ADVLOR{k}{\II^*(\nu[j-1],\nu[j]), \RR}$.
  
  Now, fix a $j\in[2,n]$. After the protocol execution $(\II^*,\RR)$,
  $\RR$ flipping the coin $b\otr[1,2]$, $\II^*$ must guess the value
  of $b$, based on his private input $(\mu[1],\mu[2])$, his private
  output $\mu[b]_1$, and the protocol view.  Since
  $\nu[j-1]_i=\nu[j]_i$ for $i\neq j$, this is equivalent to guessing
  whether $\nu[j-2+b]_j= \nu[j-1]_j$ or $\nu[j-2+b]_j= \nu[j]_b$.
  Clearly, his success is maximized here when $\nu[j-1]_j\neq
  \nu[j]_j$.  Next, $\II^*$'s view consists of $(A,B,C;\Set{(w_j,
    y_j)})$, where $(w_j,y_j) \ot (g^{r_j} A^{s_j}, g^{\mu_j}
  h^{B^{r_j} \cdot (C \cdot g^{j - 1})^{s_j}})$ for $A$, $B$ and $C$
  chosen by himself.  Since $\II^*$ is unbounded, he can find the
  value of $\alpha \neq 0$, and therefore he knows that $(w_j,y_j)=
  (g^{r_j+as_j},g^{\mu_j+\alpha B^{r_j}(C\cdot g^{j-1})^{s_j}})$.
  Since $r_j$ and $s_j$ are randomly chosen by a honest $\RR$, then
  the elements $w_j$ look completely random to $\II^*$, and do not
  help in guessing the value of $\mu_j$. He also cannot use any
  information in $(w_j,y_j)$, $j\neq j$, since these values do not
  depend on $\mu_j$.
  
  Thus, to guess the value $\nu[j-2+b]_j$, he must find a bias in the
  value $a B^{r_j}(C g^{j-1})^{s_j} = \alpha g^{b r_j+(ab+j-\sigma)
    s_j} \mod{q}$.  Note that $x\defeq \alpha g^{b r_j + (a b + j -
    \sigma) s_j}$ is a random element of $\ZZ_p^*$ due to the choice
  of $r_j$ and $s_j$, unless $b=ab+j-\sigma=0$. The latter will
  automatically hold if $i = \sigma$, but only with a negligible
  probability otherwise. Thus, we can assume that $x$ is chosen
  randomly from $\ZZ_p^*$. Guessing $\mu_j \in \ZZ_m$ from $y_j$ is
  equivalent to guessing the value $(x \mod{q}) \mod{m}$. Denote
  $e\defeq \floor{q/m}$.  Since $q \mid (p - 1)$ then $x \mod{q}$ is a
  random element of $\ZZ_q$, and $\sharp \Set{x: x\mod{q}\mod{m}=j}
  \in e+c$, where $c \in \Set{ 0, 1}$ is $1$ iff $j < d$. Thus the
  best strategy of $\II^*$ is to guess that $x$ is equivalent to some
  element $j< d$, and equivalently, that $\nu[j-2+b]_j\mod{m}\geq d$.
  He will achieve this by choosing exactly one of the two element
  $\nu[j-1]_b$ and $\nu[j-1]_b$ to have residue modulo $m$ that is
  less than $d$.  Then he will succeed with probability $e/q+1/q$
  which gives him an advantage $e/q+1/q-1/m=(m-d)/q\leq m/q$ over
  random guessing the bit $b$.  \qed
\end{proof}

Security in the left-or-right sense is both necessary and sufficient
for our purposes. Namely, in the RRT-W protocol
(Sect.~\ref{standard}), the interviewer $\II^*$ knows that the input
is --- up to the permutation of indices --- one of the two values. For
small $n$, the number of permutations is small, and thus with a high
probability $\II^*$ can guess that $\mu$ is one of the two, known for
him, Boolean vectors. Without security in the left-or-right sense, he
would be able to guess which of the two vectors is currently used, and
thus to find the type of the respondent. On the other hand, if the
oblivious transfer protocol is secure in the left-or-right sense,
$\II^*$ cannot predict the Hamming weight $w_h(\mu)=\sharp
\Set{i:\mu_1=1}$ of $\RR$'s input.

\section{Detailed Quantum CRRT}
\label{quantumdetails}

\subsection{Background on Quantum Information}\label{sec:qback}

In this section, we describe the basic notions of quantum information
needed to understand the quantum protocol and the analysis of its
simplified version in section \ref{standard}.

For a more detailed introduction to quantum information, we refer to
book by Nielsen and Chuang \cite{book:NielsenC:QC}.  A \emph{qubit} is
the basic unit of quantum information, similar to a bit in the
conventional (classical) computing.  A qubit has two basis states that
are denoted by $\ket{0}$ and $\ket{1}$.  \comment{They correspond to
  conventional $0$ and $1$.}  A general state of a qubit is
$\alpha\ket{0}+\beta\ket{1}$, with $\alpha$, $\beta$ being complex
numbers with $|\alpha|^2+|\beta|^2=1$.

We can perform two types of operations on quantum bits: unitary
transformations and measurements.  The simplest \emph{measurement} of
of a qubit $\alpha\ket{0}+\beta\ket{1}$ is in the \emph{computational
  basis} that gives the result $0$ with probability $|\alpha|^2$ and
$1$ with probability $|\beta|^2$. The state of the qubit then becomes
$\ket{0}$ or $\ket{1}$. Therefore, repeating the measurement gives the
same outcome. As long as we only consider this one type of
measurement, the state $\alpha\ket{0}+\beta\ket{1}$ behaves similarly
to a probabilistic state that has been prepared as $0$ with
probability $|\alpha|^2$ and $1$ with probability $|\beta|^2$.  This
analogy disappears, though, when we consider other transformations.
\emph{A unitary transformation} is a linear transformation on the
two-dimensional space of all $\alpha\ket{0}+\beta\ket{1}$ that
preserves the vector norm.  Two examples of unitary transformations
are the identity
$I(\alpha\ket{0}+\beta\ket{1})=\alpha\ket{0}+\beta\ket{1}$ and the bit
flip $X(\alpha\ket{0}+\beta\ket{1})=\alpha\ket{1}+\beta\ket{0}$.
\comment{ and sign flip
  $Z(\alpha\ket{0}+\beta\ket{1})=\alpha\ket{0}-\beta\ket{1}$.  To
  specify a unitary transformation, it suffices to specify $U\ket{0}$
  and $U\ket{1}$.  Then, by linearity, $U(\alpha\ket{0}+\beta\ket{1})=
  \alpha U\ket{0}+\beta U\ket{1}$.  For a transformation $U$ to be
  unitary, it is necessary and sufficient that $U\ket{0}$ and
  $U\ket{1}$ are orthogonal.} A \emph{general von Neumann measurement}
on a qubit~$\ket{\Psi}$ is specified by two orthogonal vectors
$\ket{\Phi_0}$ and $\ket{\Phi_1}$.  The outcome is either $0$ or~$1$;
the probability of outcome~$i$ is equal to the squared inner product
of $\ket{\Psi}$ and $\ket{\Phi_i}$.  The state of the qubit becomes
$\ket{\Phi_i}$. This measurement can be reduced to the measurement in
the computational basis as follows. We take a unitary $U$ that maps
$\ket{\Phi_0}$ to $\ket{0}$ and $\ket{\Phi_1}$ to $\ket{1}$. We apply
$U$ to the state $\ket{\Psi}$ that we want to measure.  Then, we
measure the resulting state in the computational basis and apply
$U^{-1}$.

\PAR{Distinguishability.}
Assume someone prepares two states $\ket{\Phi_0}$ and $\ket{\Phi_1}$,
flips a fair coin $i\otr[0,1]$, and sends $\ket{\Phi_i}$ it to us.  We
would like to guess $i$ by measuring the state.  We measure our
success by the probability that our guess $j\in\Set{0, 1}$ coincides
with $i$.  If $\ket{\Phi_0}$ and $\ket{\Phi_1}$ are orthogonal, a von
Neumann measurement in $\ket{\Phi_0}$, $\ket{\Phi_1}$ basis tells $i$
with certainty.  For non-orthogonal states, no measurement gives $i$
with certainty.

\begin{fact}\cite{book:NielsenC:QC}\label{TTheoremPure}
  The maximum success probability with what we can distinguish
  $\ket{\Phi_0}$ from $\ket{\Phi_1}$ is
  $\frac{1}{2}+\frac{\sin\beta}{2}$, $\beta$ being the angle between
  $\ket{\Phi_0}$ and $\ket{\Phi_1}$.
\end{fact}

The above definitions are sufficient to understand the protocol and
the analysis of simplified version in section \ref{standard}.  For the
full security proof, more advanced notions like \emph{density
  matrices} are needed, which are described in
Sect.~\ref{sec:density}.

\subsection{Density Matrices}\label{sec:density}

To prove the security of protocol~\ref{prot:qrrt}, we need the more
advanced formalism of \emph{density matrices}.
We interpret $\ket{\psi}=\alpha\ket{0}+\beta\ket{1}$ as a column
vector $(\alpha,\beta)^T$. Let $\bra{\psi}$ denote a row vector
$(\alpha^* \beta^*)$, with~$*$ being the complex conjugation operator.
Then, the density matrix of $\ket{\psi}$ is
\[
\ket{\psi}\bra{\psi}=
\begin{pmatrix} 
  \alpha \\
  \beta
\end{pmatrix}
(\alpha^* \beta^* ) =
\begin{pmatrix}
  \alpha\alpha^* & \alpha\beta^* \\
  \beta\alpha^* & \beta\beta^*
\end{pmatrix}\enspace.
\]

Next, assume that we generate a classical random variable that is $i$
with probability $p_i$ and then prepare a quantum state $\ket{\psi_i}$
dependent on $i$.  This creates a \emph{mixed} quantum state.  It can
be also described by a density matrix $\rho=\sum_{i}
p_i\ket{\psi_i}\bra{\psi_i}$.  If we measure a mixed state with a
density matrix $\rho$ in a basis $\ket{\Phi_0}$, $\ket{\Phi_1}$, the
probability of getting outcome $i$ is $\lbra \Phi_i | \rho |
\Phi_i\rket$ (i.e., we multiply the density matrix with the row vector
$\bra{\Phi_i}$ on the left and the column vector $\ket{\Phi_i}$ on the
right and get a number which is the probability).  The following is a
counterpart of Fact~\ref{TTheoremPure} for mixed states.

\begin{fact}\cite{book:NielsenC:QC}\label{TTheorem}
  The maximum success probability with which we can distinguish
  $\rho_0$ from $\rho_1$ is
  $\frac{1}{2}+\frac{\|\rho_0-\rho_1\|_t}{4}$, where $\|A\|_t$ is the
  trace norm of $A$ (the trace (sum of diagonal entries) of matrix
  $\sqrt{A^\top A}$).
\end{fact}

\subsection{Security Proofs for Protocol~\ref{prot:qrrt}}

\PAR{Security against Malicious Interviewer.}
\begin{theorem}\label{ISecurity}
  If a strategy for dishonest $\II^*$ leads to being caught with
  probability at most $\varepsilon$, $\II^*$ can learn $r$ correctly
  with probability at most
  $\prcorrect+\frac{2}{2\prcorrect-1}\varepsilon$.
\end{theorem}

The security of this type (cheating is possible but not without risk
of being detected) is common to many quantum protocols, for example
quantum bit commitment \cite{STOC2000:AharonovTVY} or coin flipping
\cite{PRL2002:SpekkensR}.  We note that our security guarantee is
stronger than one achieved in \cite{STOC2000:AharonovTVY}.  Namely, in
the bit commitment protocol of \cite{STOC2000:AharonovTVY}, a
dishonest party can successfully cheat with probability $\varepsilon$
so that the probability of being detected is just $O(\varepsilon^2)$.

\begin{proof}[Theorem~\ref{ISecurity}]
  Assume that we are given a strategy for dishonest $\II^*$.  First,
  notice that if we reverse the roles of $\ket{0}$ and $\ket{1}$
  everywhere in this strategy, both the probability of passing the
  test and the probability of learning $t$ correctly remain the same.
  Therefore, we can assume that the protocol is symmetric w.r.t.\ 
  switching $\ket{0}$ and $\ket{1}$.
  
  Consider the state of the first quantum bit sent by $\II^*$.  In the
  general case, $\II^*$ can send probabilistic combinations of various
  quantum states. Therefore, the first quantum bit can be in a mixed
  state with some density matrix
  \[ 
  \rho=
  \begin{pmatrix} 
    a & \alpha+\beta i \\
    \alpha-\beta i & b
  \end{pmatrix}\enspace.
  \]
  Since the strategy is symmetric w.r.t.\ switching $\ket{0}$ and
  $\ket{1}$, $\rho$ must be also symmetric in the same sense, implying
  that $a=b=1/2$ and $\beta=0$.  Thus,
  \[ 
  \rho=
  \begin{pmatrix}
    1/2 & \alpha \\
    \alpha & 1/2
  \end{pmatrix}\enspace. 
  \]
  If $\II$ is honest, $\alpha=\sqrt{\prcorrect(1-\prcorrect)}$.
  Theorem~\ref{ISecurity} follows from the following two lemmas.

  \begin{lemma}\label{ICheat1}
    The probability of $\II^*$ failing the test if the first quantum
    bit is chosen for verification is at least
    $(\sqrt{\prcorrect(1-\prcorrect)}-\alpha)\sqrt{\prcorrect(1-\prcorrect)}$.
  \end{lemma}

  \begin{lemma}\label{ICheat2}
    The probability of $\II^*$ learning $t$ correctly if the first bit
    is used for protocol and the second bit used for verification is
    at most $\frac{1}{2}+\frac{\sqrt{1-4\alpha^2}}{2}$.
  \end{lemma}
  We will for a moment assume the validity of these theorems (their
  proofs are given slightly later), and will now continue with the
  proof of the theorem.
  
  Let $\varepsilon$ be the probability with which $\II^*$ allows to be
  caught.  By Lemma~\ref{ICheat1},
  $(\sqrt{\prcorrect(1-\prcorrect)}-\alpha)
  \sqrt{\prcorrect(1-\prcorrect)}\leq \varepsilon$.  Therefore,
  $\alpha\geq \sqrt{\prcorrect(1-\prcorrect)}-
  \frac{\varepsilon}{\sqrt{\prcorrect(1-\prcorrect)}}$.  By
  substituting that into Lemma~\ref{ICheat2}, we get $
  \frac{1}{2}+\frac{\sqrt{1-4\alpha^2}}{2} \leq
  \frac{1}{2}+\frac{\sqrt{1-4\prcorrect(1-\prcorrect)+8
      \varepsilon}}{2}$.  If $\II$ is honest, the probability that
  $r=t$ is $\frac{1}{2}+\frac{\sqrt{1-4\prcorrect(1-\prcorrect)}}{2}$.
  The extra advantage gained by $\II^*$ is at most
  $\frac{\sqrt{1-4\prcorrect(1-\prcorrect)+8 \varepsilon}}{2}-
  \frac{\sqrt{1-4\prcorrect(1-\prcorrect)}}{2} \leq
  \frac{2\varepsilon}{2\prcorrect-1}$ (assuming that
  $\prcorrect>1/2$).\qed
\end{proof}
  
\begin{proof}[Lemma~\ref{ICheat1}]
  When the first bit is chosen for verification, $\II^*$ either claims
  that it is $\ket{\psi_0}$ or $\ket{\psi_1}$.  By symmetry, the
  probability of each of those is 1/2.  We partition
  $\rho=\frac{1}{2}\rho_0+\frac{1}{2}\rho_1$, with $\rho_i$ being the
  part for which $\II^*$ claims that the state is $\ket{\psi_i}$.  Let
  \[ 
  \rho_0=
  \begin{pmatrix}
    a' & \alpha' \\
    \alpha' & b'
  \end{pmatrix}\enspace.
  \]
  By symmetry, $\rho_1$ should be the same with $\ket{0}$ and
  $\ket{1}$ reversed:
  \[ 
  \rho_1=
  \begin{pmatrix}
    b' & \alpha' \\
    \alpha' & a'
  \end{pmatrix}\enspace.
  \]
  Since $\rho=\frac{1}{2}\rho_0+\frac{1}{2}\rho_1$, $a'+b'=1$ and
  $\alpha'=\alpha$. Therefore, we have
  \[ 
  \rho_0=
  \begin{pmatrix}
    a' & \alpha \\
    \alpha & 1-a'
  \end{pmatrix}\enspace.
  \]
  The probability of this state passing verification as $\ket{\psi_0}$
  is
  \begin{align*}
    \lbra \Psi_0 | \rho_0 | \Psi_0\rket = &
  \begin{pmatrix}
    \sqrt{\prcorrect} & \sqrt{1-\prcorrect}
    \end{pmatrix}
    \left( \begin{array}{cc} a' & \alpha \\ \alpha & 1-a' \end{array}
    \right) \left( \begin{array}{c} \sqrt{\prcorrect} \\
        \sqrt{1-\prcorrect} \end{array} \right)\\
    =& a' \prcorrect +
    (1-a')(1-\prcorrect)+2\alpha\sqrt{\prcorrect(1-\prcorrect)}\\
    \leq& \prcorrect^2 +
    (1-\prcorrect)^2+2\alpha\sqrt{\prcorrect(1-\prcorrect)}\\
    =& (\prcorrect+(1-\prcorrect))^2 -
    (\sqrt{\prcorrect(1-\prcorrect)}-\alpha)\sqrt{\prcorrect(1-\prcorrect)}\\
    =& 1-
    (\sqrt{\prcorrect(1-\prcorrect)}-\alpha)\sqrt{\prcorrect(1-\prcorrect)}\enspace.
  \end{align*}\qed
  \end{proof}

  \begin{proof}[Lemma~\ref{ICheat2}]
    We assume that the second qubit has been prepared perfectly and
    its verification always succeeds.  (If $\II^*$ cheated in
    preparing the second qubit as well, this only decreases the
    probability of success for $\II^*$ and the claim that we prove
    remains valid.)
  
    After the test is passed on the second qubit, $\RR$ has the first
    qubit in the mixed state $\rho$. The mixed state $\rho$ is the
    same as one obtained by taking
    $\frac{1}{\sqrt{2}}\ket{0}+\frac{1}{\sqrt{2}}\ket{1}$ with
    probability $2\alpha$ and $\ket{0}$, $\ket{1}$ with probabilities
    $\frac{1}{2}-\alpha$ each.  Therefore, the joint state of $\II^*$
    and $\RR$ is equivalent to
    $\ket{\psi_{(\RR,\II^*)}}=\sqrt{\frac{1}{2}-\alpha}\ket{0}_{\II^*}\ket{0}_{\RR}+
    \sqrt{\frac{1}{2}-\alpha}\ket{1}_{\II^*}\ket{1}_{\RR}+
    \sqrt{2\alpha} \ket{2}_{\II^*}
    (\frac{1}{\sqrt{2}}\ket{0}+\frac{1}{\sqrt{2}}\ket{1})_{\RR}$. If
    $\RR$'s secret bit $t=0$, he just sends his part back to $\II^*$.
    After that, $\II^*$ possesses the entire state
    $\ket{\psi_{(\RR,\II^*)}}$. Otherwise, $\RR$ flips the qubit
    before sending back and $\II^*$ gets
    $\ket{\psi'_{(\RR,\II^*)}}=\sqrt{\frac{1}{2}-\alpha}\ket{0}_{\II^*}\ket{1}_{\RR}+
    \sqrt{\frac{1}{2}-\alpha}\ket{1}_{\II^*}\ket{0}_{\RR}+
    \sqrt{2\alpha} \ket{2}_{\II^*}
    (\frac{1}{\sqrt{2}}\ket{0}+\frac{1}{\sqrt{2}}\ket{1})_{\RR}$. Now,
    the question is how well can $\II^*$ distinguish these two states.
    By Fact~\ref{TTheoremPure}, the best probability with which he can
    get $t$ is
    $\frac{1}{2}+\frac{\sin\beta}{2}=\frac{1}{2}+\frac{\sqrt{1-\cos^2\beta}}{2}$
    where $\beta$ is the angle between the two states.  $\cos\beta$ is
    equal to the inner product of $\ket{\psi_{(\RR,\II^*)}}$ and
    $\ket{\psi'_{(\RR,\II^*)}}$ which is $2\alpha$ ( because the first
    two components of $\ket{\psi_{(\RR,\II^*)}}$ are orthogonal to the
    first two components of $\ket{\psi'_{(\RR,\II^*)}}$ but the third
    component is equal).\qed
\end{proof}

\PAR{Security against Malicious Respondent.}

\begin{theorem}
\label{thm:malres}
Let $\prcorrect<\frac{1}{2}+\frac{ \sqrt{3} }{4}=0.933...$.  If $\II$
is honest, $\RR^*$ cannot achieve $t=0$ (or $t=1$) with probability
more than $\pradv\leq \frac{1}{2}+
\sqrt{\sqrt{4\prcorrect-4\prcorrect^2}-(4\prcorrect-4\prcorrect^2)}$.
\end{theorem}

The probability $\pradv$ remains less than 1 for all
$\prcorrect<0.933...$.  Thus, our protocol offers nontrivial security
guarantees for all $\prcorrect<0.933...$.  Since the expression for
$\pradv$ is quite complicated, we also present a simple but less
precise bound.  Let $\prcorrect=\frac{1}{2}+\epsilon$.  Then,
$\pradv\leq \frac{1}{2}+\sqrt{2}\epsilon$.  Informally, this means
that no $\RR^*$ can make his vote count as more than $\sqrt{2}$ votes.
This gives a non-trivial bound on $\pradv$ for $\prcorrect<
\frac{1}{2}+\frac{1}{2\sqrt{2}}=0.853...$.

If $0.853...\leq \prcorrect\leq 0.933...$, then
$\frac{1}{2}+\sqrt{2}\epsilon\geq 1$ but $\pradv<1$ which can be seen
by evaluating the expression of theorem \ref{thm:malres} directly.

\begin{proof}
  There are four possible states that a responder can receive from an
  honest $\II$: $\ket{\psi_0}\ket{\psi_0}$,
  $\ket{\psi_0}\ket{\psi_1}$, $\ket{\psi_1}\ket{\psi_0}$,
  $\ket{\psi_1}\ket{\psi_1}$.  An honest responder then randomly
  requests to verify one of two quantum bits. A dishonest $\RR^*$ can
  measure the state and then decide to verify one of two bits based on
  the result of the measurement so that his chances of guessing the
  other state are maximized.  Without loss of generality, $\RR^*$'s
  measurement has two outcomes: $0$ and $1$ and the index $i$ that is
  sent back to $\II$ is equal to the outcome of the measurement.
  Then, we have
\[ \ket{\psi_{u_0}\psi_{u_1}}=
\alpha_{u_0u_1}\ket{0}\ket{\psi'_{u_0u_1}}+
\beta_{u_0u_1}\ket{1}\ket{\psi''_{u_0u_1}} ,\] where the first qubit
is the one being measured and $\ket{\psi'_{u_0u_1}}$
($\ket{\psi''_{u_0u_1}}$) is the rest of the quantum state that
remains with $\II$ after the measurement. By symmetry, we can assume
that $\alpha_{u_0u_1}=\beta_{u_0 u_1}=\frac{1}{\sqrt{2}}$.

Similarly to the simplified protocol in Sect.~\ref{standard}, the
probability of $\RR^*$ fixing $r=0$ (or $r=1$) is equal to the
probability that he correctly guesses $u_{1-i}$. We bound this
probability.  For brevity, assume that $\RR^*$ has requested $u_1$
from $\II$ and received $u_1=0$. Then, if $u_0=0$, his remaining state
is $\ket{\psi'_{00}}$ and, if $u_0=1$, his remaining state is
$\ket{\psi'_{10}}$. The probability with which he can guess $u_0$ is,
by Fact~\ref{TTheoremPure}, at most
$\pradv=\frac{1}{2}+\frac{\sin\beta'}{2}$ where $\beta'$ is the angle
between $\ket{\psi'_{00}}$ and $\ket{\psi'_{10}}$.  Remember that, by
analysis of Sect.~\ref{standard}, the probability of $r=t$ in the
honest case is described by similar expression
$\prcorrect=\frac{1}{2}+\frac{\sin\beta}{2}$ where $\beta$ is the
angle between $\ket{\psi_0}$ and $\ket{\psi_1}$.

Next, we express $\beta'$ by $\beta$.  Remember that $\lbra
\psi|\psi'\rket$ denotes the inner product between $\ket{\psi}$ and
$\ket{\psi'}$.  The inner product $\lbra \psi_0 | \psi_1\rket$ is
equal to $\cos\beta$. The inner product between
$\ket{\psi_0}\ket{\psi_0}$ and $\ket{\psi_1}\ket{\psi_0}$ is the same
$\cos\beta$ because the second qubit is in the same state in both
cases. This inner product is also equal to $\frac{1}{2} \lbra
\psi'_{00} |\psi'_{10}\rket+ \frac{1}{2} \lbra \psi''_{00}
|\psi''_{10}\rket$.  The first part is $\cos\beta'$, the second part
is at most 1.  Therefore, $\frac{1}{2}(\cos\beta'+1)\geq \cos\beta$
and $\cos\beta'\geq 2 \cos\beta -1$.  We have
$\sin\beta'=\sqrt{1-\cos^2\beta'}\leq \sqrt{4(\cos\beta-\cos^2
  \beta)}$ and $\pradv\leq \frac{1}{2}+\frac{\sin\beta'}{2}\leq
\frac{1}{2}+\sqrt{\cos\beta-\cos^2 \beta}$.  Remember that in the
honest protocol, the probability that $r=t$ is
$\prcorrect=\frac{1}{2}+\frac{\sin\beta}{2}$.  Therefore,
$\sin\beta=2\prcorrect-1$,
$\cos\beta=\sqrt{1-\sin^2\beta}=\sqrt{4\prcorrect-4\prcorrect^2}$ and,
by substituting this into
$\pradv\leq\frac{1}{2}+\sqrt{\cos\beta-\cos^2 \beta}$, we get the
theorem.  \qed\end{proof}

To show the $\pradv\leq \frac{1}{2}+\sqrt{2}\epsilon$ upper bound, it
suffices to show $\sqrt{\cos\beta-\cos^2\beta}\leq \sqrt{2}\epsilon$.
Since $\epsilon=\frac{\sin\beta}{2}$, this follows from
\[ \frac{\sqrt{\cos\beta-\cos^2\beta}}{(\sin\beta)/2} =
\frac{2\sqrt{\cos\beta-\cos^2\beta}}{\sqrt{1-\cos^2\beta}} =
\frac{2\sqrt{\cos\beta}}{\sqrt{1+\cos\beta}} \leq
\frac{2\sqrt{\cos\beta}}{\sqrt{2\cos\beta}}=\sqrt{2} \]

\end{document}